# Evaluation of reference region modelling techniques in quantitative dynamic contrast-enhanced magnetic resonance imaging of breast cancer

Christian Notte[1,2], Zaki Ahmed[3], and Ives R. Levesque [1,2,4,*]

[1] Medical Physics Unit, McGill University, Montréal, Canada

[2] Gerald Bronfman Department of Oncology, McGill University, Montréal, Canada

[3] Corewell Health William Beaumont University Hospital, Royal Oak, USA

[4] Cancer Research Program, Research Institute of the McGill University Health Centre, Montréal, Canada

* Corresponding author: Ives Levesque

ives.levesque@mcgill.ca

McGill University Health Centre – Glen Site, room DS1.9326

1001 Decarie Boulevard

Montreal QC H4A 3J1

Canada

Abstract: 236

Manuscript excluding figures and tables: 4127

## Abstract

**Purpose:** To assess the validity of reference region model (RRM) techniques for quantitative perfusion analysis of low temporal resolution DCE-MRI of breast lesions.

**Methods:** Tofts model (TM), extended Tofts model (ETM), RRM, and extended RRM (ERRM) were applied to a dataset of images from ten patients with confirmed breast cancer collected before and after one cycle of chemotherapy, with known pathologic treatment response. Quantitative maps of estimated perfusion parameters were produced using each model, and changes in median perfusion parameters before and after a cycle of chemotherapy were compared across models. Relative parameters from reference region models were made absolute using the reference region and input function tail (RRIFT) technique. Tumour region voxel count before and after treatment was also recorded.

**Results:** RRMs produced similar absolute quantitative analysis compared to AIF-driven models (TM and ETM), with scaling differences in $K^{trans}$ parameter estimates and disagreement in $v_p$ parameter estimates between ETM and ERRM. RRMs and AIF-driven models were equally capable of separating complete responders based on change in median $K^{trans}$ parameter, while change in tumour voxel count was not predictive of treatment response. Both RRMs were also able to fully separate complete responders using relative $K^{trans}$ parameter estimates, with no AIF information.

**Conclusions:** RRMs show similar performance as standard models in quantitative analysis of breast lesions, with reduced reliance on the AIF. Results imply that RRMs may be a suitable alternative at low temporal resolutions to provide reliable, patient-specific quantitative analysis.

## Introduction

Dynamic contrast-enhanced magnetic resonance imaging (DCE-MRI) is a clinically adopted technique which allows for the non-invasive analysis of tissue perfusion. This is accomplished through the collection of multiple $T_1$-weighted images during the passage of a paramagnetic tracer through the tissue of interest[1,2]. The application of DCE-MRI in cancer imaging is intuitive, due to the tortuous, leaky vasculature that accompanies growing malignancies[3]. This is especially true of breast tumours, which are particularly angiogenic[4]. In addition, while changes in tumour size (manually-defined volume or diameter) are the most common measure for assessing treatment response, it has been noted that these changes are subjective and often manifest after changes in tumour physiology, particularly perfusion[5,6]. Thus, researchers have sought to apply DCE-MRI to breast cancer imaging, showing value in diagnosis[7], grading[8], and treatment response monitoring[5–7,9].

While time-course imaging can be analyzed qualitatively to provide general information about perfusion in pathology[10], quantitative analysis using pharmacokinetic models produces richer information on a voxel-wise basis. The most common pharmacokinetic model used in DCE-MRI is the Tofts model (TM), favoured for its stability when faced with poor data quality[11–13]. The TM provides two independent perfusion parameters: the volume transfer constant $K^{trans}$, which estimates a combination of tissue flow and permeability, and $v_e$, the fractional volume of the extravascular-extracellular space (EES) that is accessible to the tracer. A third parameter is frequently used, the flux rate constant between the EES back to the plasma ($k_{ep}$), which is equal to the ratio $K^{trans}/v_e$[14]. The TM assumes that within the voxel the blood plasma volume ($v_p$) is zero. To address this, the Extended Tofts model (ETM), frequently termed the standard model in literature, was derived[14]. This development was especially important for the application of cancer imaging where the malignant vasculature is often non-negligible[15,16], though the extra fitting term produces some extra instability in the model[17].

Quantitative DCE-MRI has shown great promise; however, its widespread usage remains limited due to technical constraints[4]. This is largely the case due to the added demand of temporal resolution on top of the typical image quality requirements. Indeed, for nearly all quantitative DCE-MRI, an arterial input function (AIF) is required and accurate sampling of its rapid temporal dynamics is very important for accuracy of quantitative analysis[18]. Since the most convenient way of sampling the AIF is from the image series itself, a high temporal resolution (≤ 1 second) is required to accurately capture the rapid signal fluctuations that comprise the first-pass dynamics of the AIF[19]. This causes strain in balancing other image quality requirements such as contrast-to-noise ratio (CNR) and high spatial resolution, both of which are important for visualizing small lesions and for accurate quantification. In breast MRI, it is also often desirable to capture both breasts within the field of view (FOV)[2], adding another constraint to an already demanding protocol. Due to these challenges, investigators often compromise by choosing between a high temporal resolution for accurate perfusion quantification, or a high spatial resolution and instead use qualitative/semi-quantitative methods[7].

To ease the imaging burden, many researchers have sought to reduce the temporal requirements stipulated by AIF sampling. The most common technique has been to replace the patient-specific AIF with a literature-based AIF, either by directly using an averaged curve or an analytic functional form derived from prior patient datasets[20–22]. While some investigators have argued population-derived AIFs may in fact be beneficial[20,23], others argue that they miss critical patient-specific variations[21,24–26].

Another approach that seeks to retain patient-specific information is the reference-region model (RRM), which has seen use both in animal[9,27,28] and human studies[8,29]. The RRM is a reformulation of the TM, which uses the time course of a separate healthy tissue (termed the reference region) that receives the same vascular supply as the tissue of interest, as a surrogate to the AIF[30]. Since first-pass tracer kinetics in healthy tissue are much slower than in feeding arteries, this avoids the rapid temporal dynamics of the AIF that constrain the imaging protocol, while still providing patient-specific information. The RRM has been shown to lead to overall improved parameter estimate accuracy at lower temporal resolutions[7]. A recent study comparing the TM and RRM found that the relative $K^{trans}$ estimates of the RRM provided the best indication of treatment effects after injection of a vascular-disrupting agent in rodents[31]. The largest drawbacks of the RRM has been its lack of inclusion of a plasma volume compartment, greater variability due to the introduction of extra fitting parameters, as well as the need to assume reference region perfusion parameters for absolute quantification[1]. These drawbacks have been recently addressed through the development of the extended reference region model (ERRM)[32] and the reference region and input function tail (RRIFT) method[33], respectively, which have been tested in simulation and a dataset from patients with confirmed glioblastoma. Additionally, constrained versions of these models have been developed which use a two-step fitting approach to reduce fitting dimensionality and improve parameter estimates[34].

In this work, we compared the performance of four RRMs with the TM and ETM in breast cancer perfusion imaging. We hypothesized that, in a low temporal resolution dataset, the RRMs would provide improved quantitative analysis of the breast lesions over standard single-compartment models, since temporal dynamics of muscle are slower than the AIF. This was assessed by comparing quantitative maps as well as the distributions of parameter estimates made by each model. Further, the inclusion of imaging before and after the first cycle of chemotherapy, alongside pathologic response data allowed for the assessment of the predictive ability of each model, providing a measure of clinical usefulness.

## Methods

Six pharmacokinetic models were applied to the data: the TM, ETM, LRRM, ERRM, CLRRM, and CERRM. The LRRM here is a reformulation of the RRM[30], that has been linearized to aid with least-squares fitting[35]. The TM and ETM have been linearized as well, though in this work this is omitted from the model name[36]. The ERRM is an extended version of the RRM, analogous to the extended Tofts model, that accounts for non-zero plasma volume within the tumour region[32]. In the LRRM and ERRM, the reference region is assumed to have zero plasma volume, which is a reasonable assumption in healthy muscle[37,38]. The CLRRM and CERRM models are the 'constrained' versions of the two reference region models, as described previously. Each model provided voxel-wise estimates of two or three pharmacokinetic parameters.

The six models were evaluated in-vivo using DCE-MRI data from a longitudinal study of response to neoadjuvant chemotherapy, obtained from The Cancer Imaging Archive[39]. This dataset contains 10 patients with confirmed breast cancer who underwent imaging at a single site including a $T_1$-weighted DCE-MRI time series with a temporal resolution of 16–18 s. Imaging was conducted both before and after the first cycle of chemotherapy. Patient response was also specified in the data (exact definitions included in the related publication), with three patients being indicated as complete responders (CR) and the remaining seven being indicated as non-complete responders (nCR)[40]. The dataset included contours of the breast lesion drawn by experienced radiologists and a fixed tumour $T_1$ value. Due to the low temporal resolution, an AIF could not be extracted by contouring a feeding vessel within the DCE time course. Instead, the AIF used was a population-average from the AIFs acquired in a previous, high resolution (2 s), breast DCE-MRI study with the same contrast agent injection protocol[4,41]. In our analysis, this high-resolution representative AIF was shifted to align the bolus arrival timing with the enhancement of the tumour region, under the logic that the bolus injection protocol is fixed per imaging session. Once shifted, the AIF was subsequently downsampled to match the temporal resolution of the imaging time series by taking the median value of the AIF at each time point. Additional information regarding the dataset, including data acquisition and analysis, can be found in the related publication[40].

From the provided contours, tumour volume was estimated by calculating the number of tumour voxels; subsequently, the percentage change in volume was calculated between imaging sessions. Reference region contours of nearby muscles were drawn by one of the authors (C.N.). The ipsilateral pectoralis major was the primary reference region. The contralateral pectoralis major (patients 4, and 6) and ipsilateral serratus anterior (patients 3, 5, and 10) were used in cases of lesion infiltration or poorly controlled chest motion, respectively. All pre-processing, model fitting, and analysis was done using in-house functions in MATLAB (The MathWorks, Natick, MA, USA). Much of the code used in this work developed on existing code from previous projects[32–34]. Prior to fitting, percentile-based outlier rejection (removal of the largest and smallest 5%) was conducted on the $R_1$ values of the tissue of interest, and any non-enhancing voxels were set to zero. The AIF-tail, used by the RRIFT technique to make the relative perfusion estimates by the RRMs absolute, was taken as the provided AIF starting from the first datapoint after the second minute until the end of acquisition.

Quantitative maps of the tumour region of each patient were generated using the voxel-wise model fit outputs. The root mean squared error (RMSE) was also mapped to assess fit quality. Box-and-whisker plots were produced to summarize trends in the $K^{trans}$ and $v_p$ estimate distributions across the entire dataset. 2D voxel density histograms were also generated from the same output to assess the agreement between models. To quantify the agreement, the concordance correlation coefficient (CCC) and Pearson correlation coefficient ($\rho$) were calculated, which are included in their entirety in the Supplementary Information. Briefly, the Pearson correlation coefficient measures the linearity between two sets of data, while the CCC specifically also measures how close the linear relationship is to identity, giving a sense of how numerically similar the data is. Finally, changes after treatment were investigated by calculating the median of each pharmacokinetic parameter within the tumour region for each imaging session and calculating the percentage change of these values between imaging sessions.

## Results

According to the provided contours, tumour sizes largely decreased between visits one and two, with the exception of patients 3 and 9, whose tumour size increased by 87% and 210% respectively. These findings are summarized in Table 1. The complete responders (CR) did not show any distinct trends in volume change that separate them from the non-complete responders (nCR).

**Table 1: Late-enhancing gross tumour volume voxel count, for each patient, on each day, as well as their percentage change. All patients showed a slight to large decrease in number of voxels within the tumour contour region between days 1 and 2 (far right column), except for patients 3 and 9. Complete responders are indicated by a black outline.**

| Patient Number | Day 1 voxel count | Day 2 voxel count | Percentage Change |
|---|---|---|---|
| **1** | 12683 | 8938 | -29.5 |
| **2** | 2143 | 985 | -54.0 |
| **3** | 738 | 1382 | 87.3 |
| **4** | 47470 | 28525 | -39.9 |
| **5** | 3682 | 1767 | -52.0 |
| **6** | 6130 | 2163 | -64.7 |
| **7** | 1058 | 679 | -35.8 |
| **8** | 3997 | 2693 | -32.6 |
| **9** | 3410 | 10577 | 210.2 |
| **10** | 5147 | 1658 | -67.8 |

All four models produced maps of perfusion parameters with spatial patterns that are physiologically plausible, with some variation between them. Example maps from one patient (patient BC01, day 1, slice 56) are shown in Figure 1. While the spatial features in $K^{trans}$ estimates are similar across models, the two AIF-driven models appeared to exhibit overall lower $K^{trans}$ estimates compared to the RRMs. Qualitative differences in spatial features and heterogeneity of $v_p$ estimates were observed between the ETM and CERRM. The ETM, TM and CLRRM show qualitatively similar RMSE maps with the ETM having the lowest overall RMSE of the three. The CERRM shows differing behaviour, with some voxels on the edges of the gross tumour volume having a higher RMSE than all other models, while overall the median RMSE was the lowest of all models tested. Some heterogeneity was observed among all maps, though this is expected as malignant tumours are most often inherently heterogenous[42,43]. Similar RMSE behaviour was seen across all patients, shown for both pre- and post- initial treatment in Figure S.1.

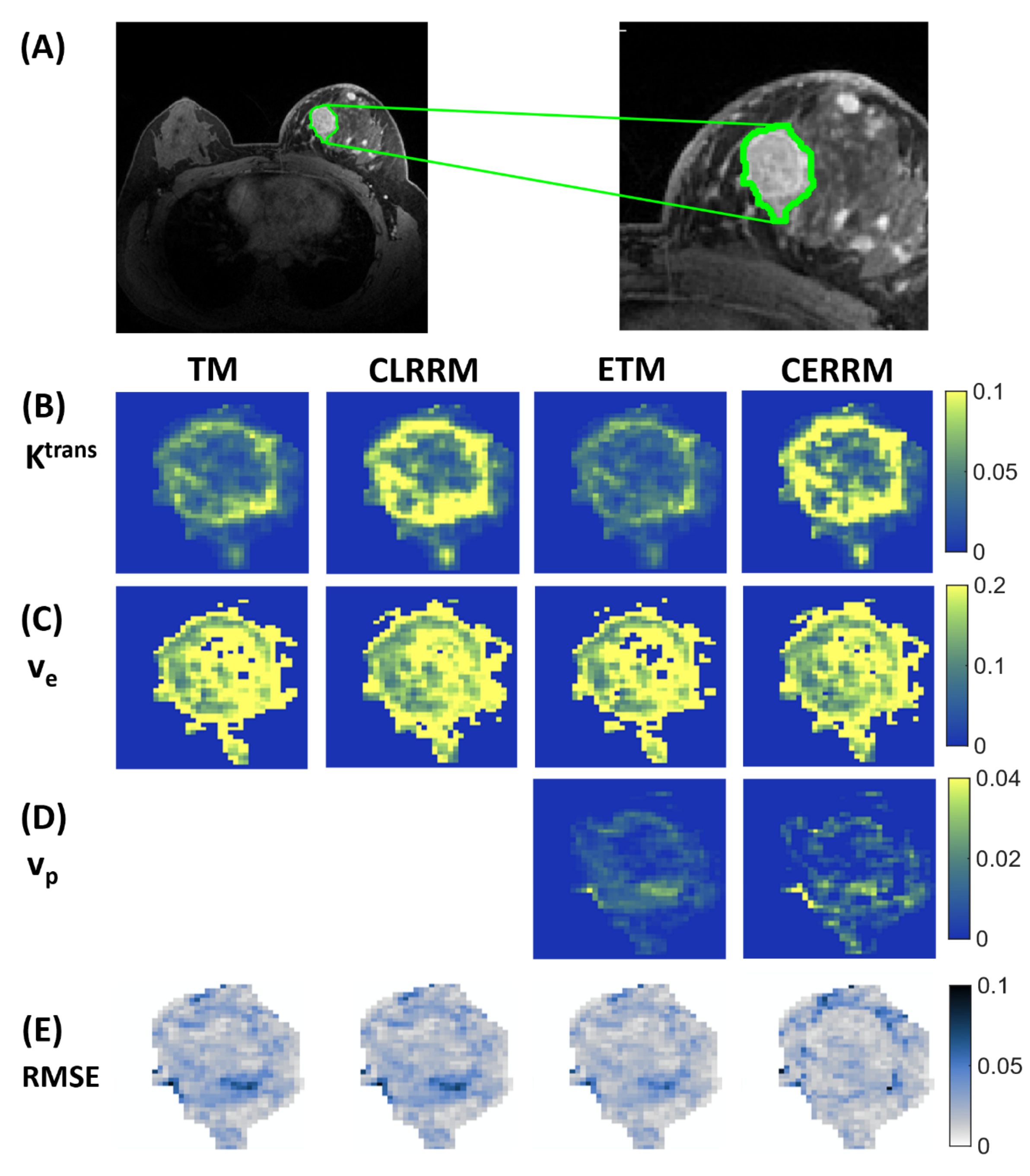

**Figure 1: Representative slice of one patient on day 1 (patient 1, day 1, slice 56). A full slice and a cropped portion of one of the DCE magnitude images is included as an example (A), with the tumour region indicated by the dotted green contour. (B)-(D) are the corresponding pharmacokinetic maps generated using each model, for Ktrans, kep and vp, respectively. (E) are the corresponding maps of root mean square error per voxel, for each model.**

In comparing voxel-wise model estimates across the entire tumour region, all models showed moderate agreement in $K^{trans}$ values (Figure 2A-F, left column); this is visualized by the high voxel density that occurs along the diagonal for very low $K^{trans}$ values, beyond which it begins to diverge. For most comparisons this divergence was characterized by an overall increased spread of values, sometimes shifting above or below the diagonal, indicating possible biases between models. The ETM and TM appeared to show the closest agreement in voxel estimates, corroborated by the high CCC and $\rho$ values. Comparisons between either RRM and the TM, and between the two RRMs themselves, showed modest agreement, while the poorest agreement was observed between the CERRM and ETM. Certain model comparisons (Figure 2C-E) showed a unique pattern in which the divergence at increasing $K^{trans}$ values was comprised of discrete patient-specific clusters with slopes greater/smaller than unity. This is especially evident in the comparisons between the CLRRM and TM, but also is true of the CLRRM and ETM, and the CLRRM and CERRM comparisons. The $K^{trans}$ estimates of CERRM and TM exhibit the least amount of divergence at higher $K^{trans}$ values, though the estimates curve above the diagonal, resulting in relatively high $\rho$ but poor CCC. The comparison between the two RRM $K^{trans}$ estimates showed modest agreement, here showing two distinct clusters of patients (Figure 2E); patients 4, 8, and 9 comprising the lower cluster of points, and the remainder comprising the upper cluster, each indicated with dotted grey lines.

Estimates of $v_e$ demonstrated strong linearity across all model comparisons, supported by the high correlation coefficients and the concentration of voxel density along the diagonal. The largest disagreement was seen in comparisons between CERRM and either AIF-driven model. Like the $K^{trans}$ estimates, disagreement was greater at higher $v_e$ values. Finally, the $v_p$ comparison between CERRM and ETM (Figure 2A, right-most column) showed a large degree of dispersion, reiterating the disagreement seen in the quantitative maps (Figure 1D).

All models showed approximate agreement in median tumour $K^{trans}$ values across patients, with ETM generally providing the lowest median estimates and CERRM providing the highest. This is visualized in Figure 3. The median $v_p$ estimates by CERRM tended to fluctuate greatly across patients, whereas the ETM values were more stable. Similar trends were observed in the data from the second imaging session, provided in the supplementary materials (Figure S.2). For the first visit imaging, CERRM exhibited the greatest interquartile range of $K^{trans}$ values, followed by CLRRM and TM, then by ETM with the tightest ranges. Notably, the interquartile ranges of the $K^{trans}$ estimates among CRs exhibited a significant reduction in both RRMs after a cycle of treatment. For $v_p$, the ETM shows smaller interquartile ranges than the CERRM.

Comparing the change in tumour perfusion characteristics between pre- and post-treatment imaging sessions, all models returned a similar range of percentage change for $K^{trans}$, with treatment-induced changes being model dependent (Figure 4). All models show the largest decrease in $K^{trans}$ to occur among CRs, which is consistent with previous findings for the AIF-reliant models[2]. Considering $\Delta k_{ep}$, CRs exhibited greater decreases in $k_{ep}$ than most nCRs, for all models, with the CERRM being the only model to entirely separate CRs from nCRs on this basis. Observing changes in the compartment fractional volumes, the AIF-based models largely show an increase in $v_e$, with only two nCRs showing a slight decrease. The RRMs instead show changes centred closer to zero, and all CRs showing a slight to moderate decrease. As for $v_p$, both the ETM and CERRM show somewhat erratic behaviour, with CRs showing sizable increase or decreases, and some nCRs showing very large increase.

The 'constrained' fitting approach was found to have a strong impact on the $\Delta K^{trans}$ separation of CRs and nCRs, as well as overall estimate stability, for both RRMs. The RRIFT step further reduced the spread of parameter estimates, though it was not strictly necessary for separation of CRs from nCRs based on $\Delta K^{trans}$. Additionally, both the constraining step and the RRIFT step substantially reduced the percentage change estimates in both $K^{trans}$ and $v_e$. These findings are summarized in Figure 5.

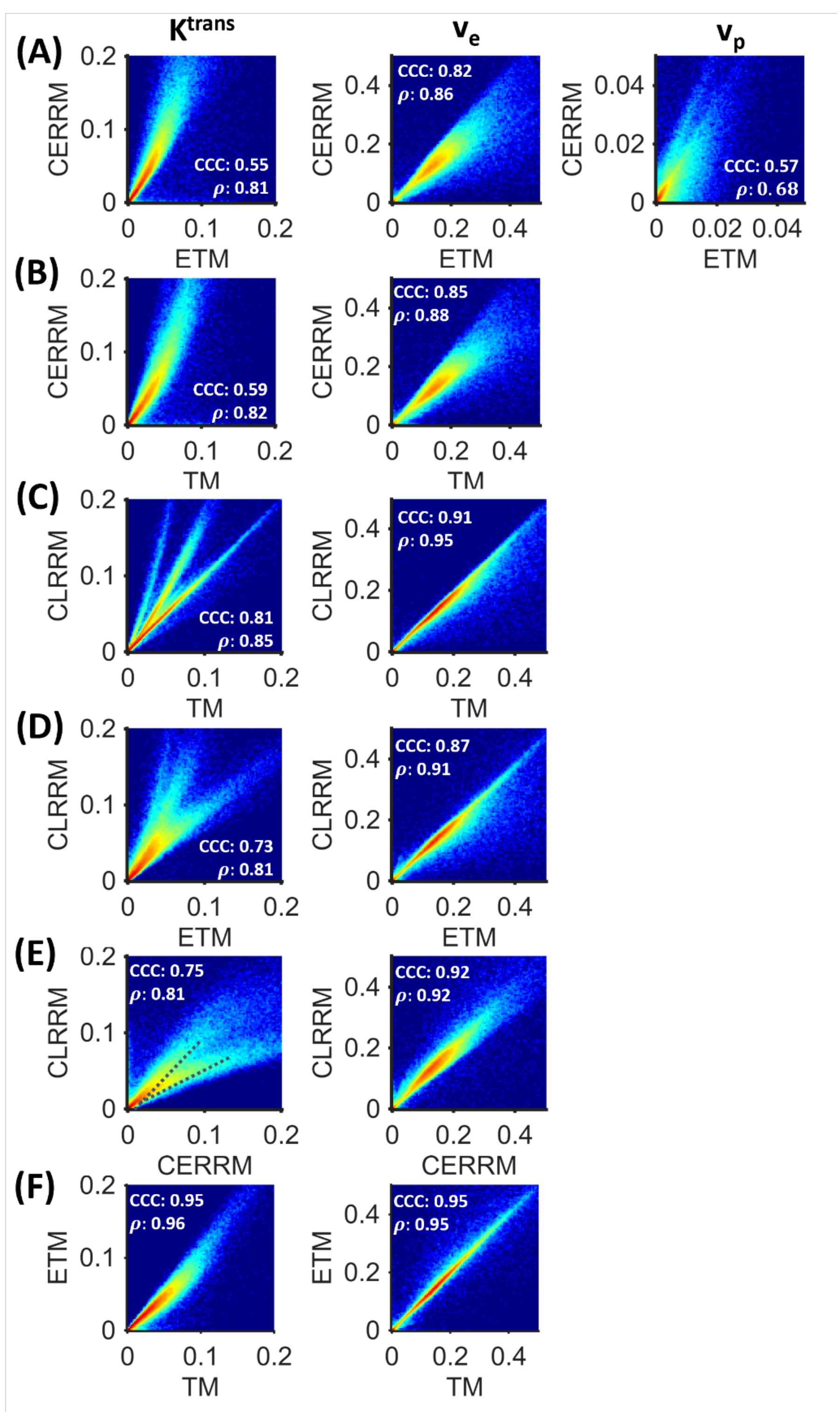

**Figure 2: 2D voxel density histograms of pharmacokinetic parameters generated between (A) CERRM and ETM, (B) CLRRM and ETM, (C) CERRM and TM, (D) CLRRM and ETM, (E) CLRRM and CERRM, and (F) ETM and TM. The concordance and Pearson correlation coefficients (CCC and $\rho$) are included to demonstrate linearity of model agreement. Note that perfect agreement between any two models on estimates of any pharmacokinetic parameter would be suggested by a diagonal line of red intercepting the origin.**

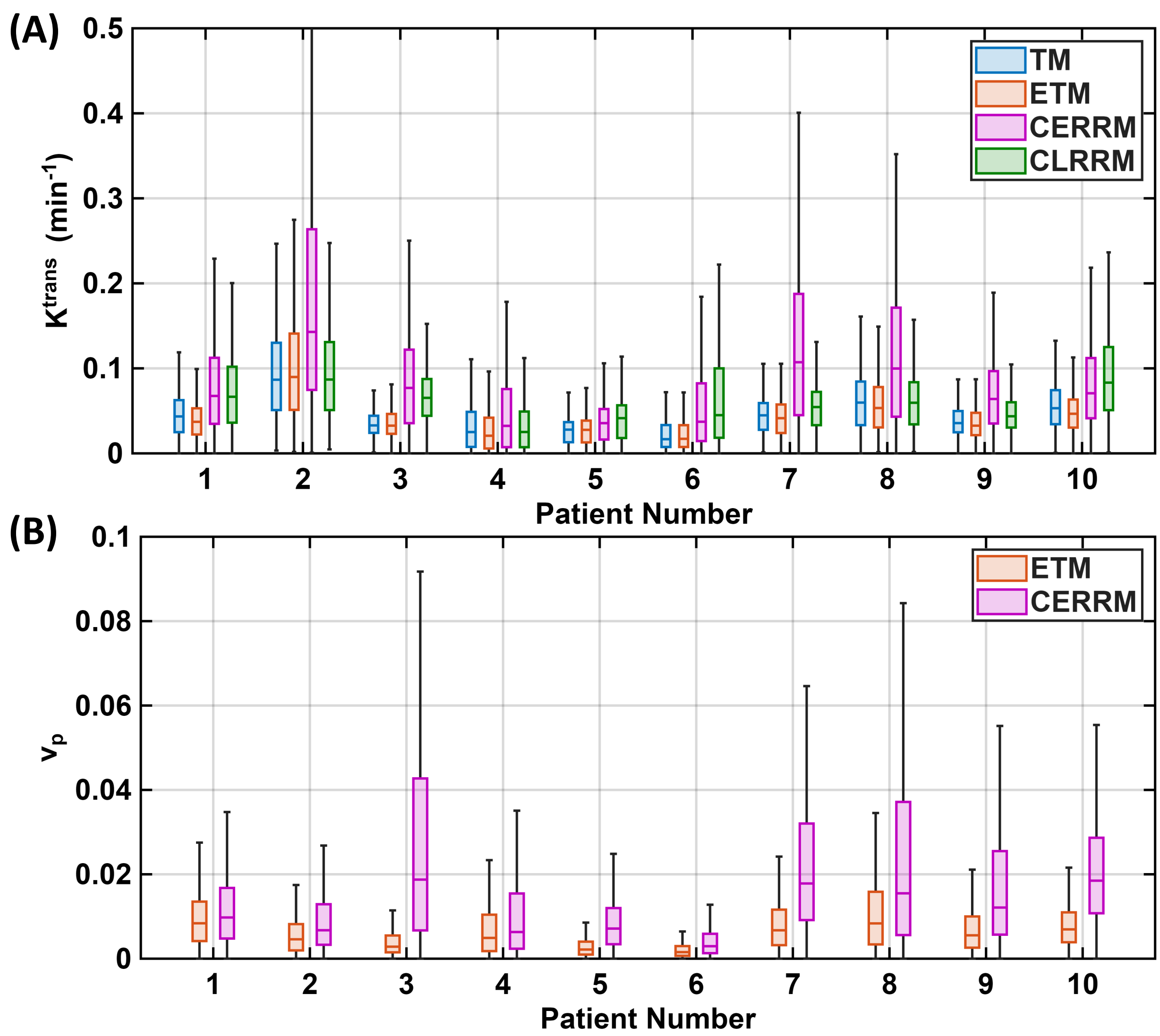

**Figure 3: Box-and-whisker plots of $K^{trans}$ (A) and $v_p$ (B) estimates obtained from each pharmacokinetic model, for each patient. The box indicates the interquartile range, with the horizontal line contained within indicating the median value. The whiskers indicate the minimum and maximum values.**

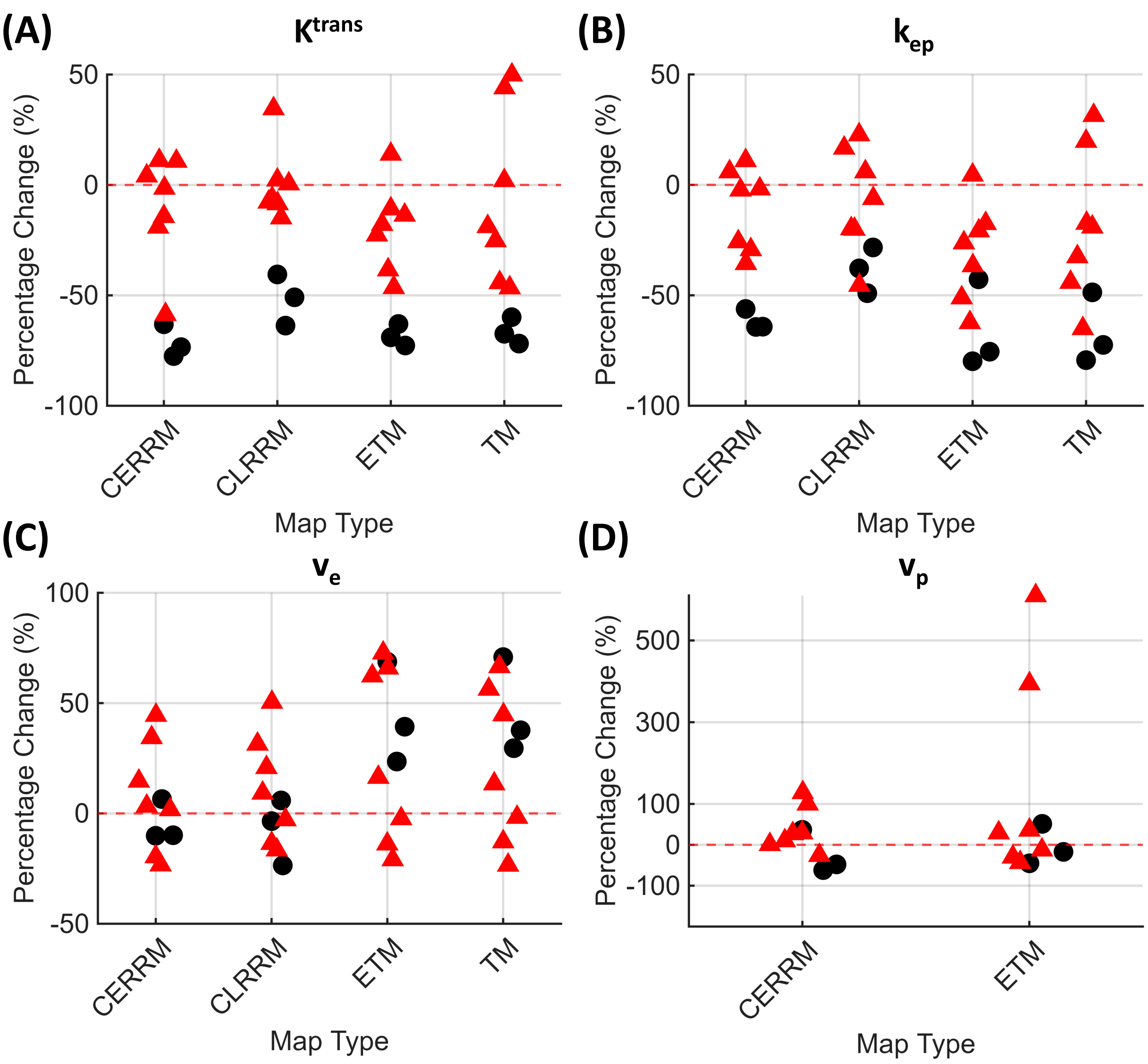

**Figure 4: Percent change in median tumour value of $K^{trans}$ (A), $k_{ep}$ (B), $v_e$ (C), and $v_p$ (D), from each of the pharmacokinetic models, before and after chemotherapy treatment. Each data point represents an individual patient with the black circles indicating patients who exhibited a complete response (CR), while red triangles are patients who showed non-complete response (nCR).**

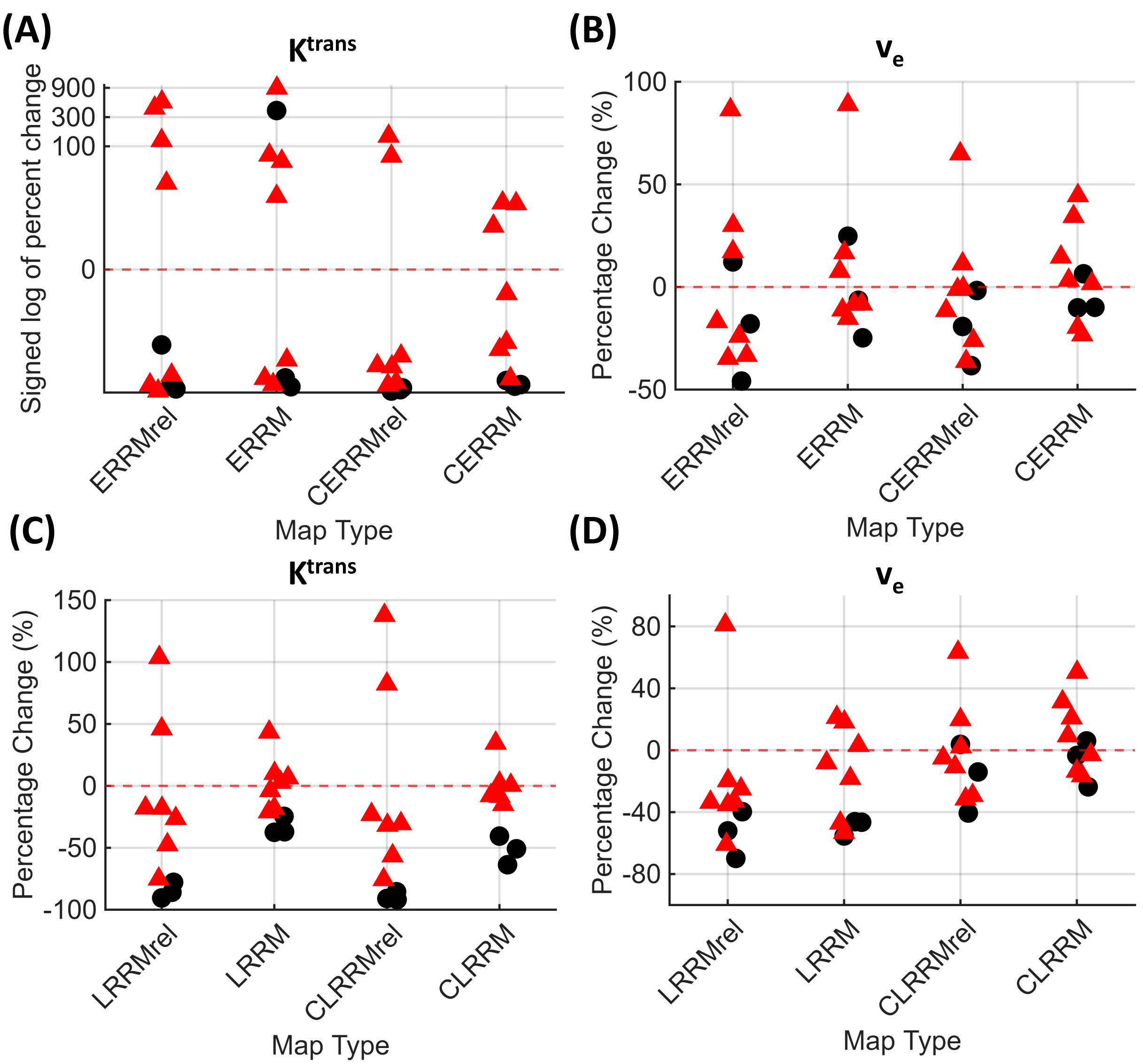

**Figure 5: Percent change in $K^{trans}$ and $v_e$ value before and after chemotherapy treatment, from extended (A and B) and non-extended (C and D) reference region models, respectively. Note the change in $K^{trans}$ estimates from the ERRM models uses a signed logarithmic y-axis, for ease of viewing both large outliers and trends between patients. Each data point represents an individual patient with the black circles indicating patients who exhibited a complete response (CR), while red triangles are patients who showed non-complete response (nCR).**

## Discussion

Overall, findings suggest that RRMs can produce quantitative analysis of breast tumours comparable to standard AIF-driven models, yielding physiologically reasonable quantitative maps with similar distributions of perfusion parameter estimates throughout each tumour volume. Systematic differences were observed between $K^{trans}$ and $v_p$ estimates. Estimates of $K^{trans}$ from each model largely appeared to differ approximately by a scaling factor; however, the curvature of the dispersion seen in the voxel density histograms implies that this is not the only difference. The reference region models generally estimate a wider range of $K^{trans}$ values. This range greatly reduced in CRs in the second visit, which, when considered along the decrease in median $K^{trans}$ seen in CRs, could be interpreted as a reduction in tumour heterogeneity and of high perfusion subregions. For the $v_p$ estimates, disagreement is much greater between the ETM and CERRM, with some shared hot-spots but otherwise far more heterogeneity in the CERRM estimates.

Using limited information from the AIF (i.e. only the AIF-tail for the absolute estimates, and no AIF at all for the relative estimates), RRMs also proved to be equally capable of separating CRs from nCRs in this dataset on the basis of larger decreases in $K^{trans}$ over the first cycle of chemotherapy, in agreement with observations from the AIF-driven models. This result is meaningful as, in the absence of a ground truth, clinical endpoints serve as a strong method of assessing the viability of different techniques. This discriminative ability was observed even among the relative estimates of the constrained models that skip the RRIFT technique. The ability to distinguish between CRs and nCRs without needing an AIF is meaningful and, in the case of this dataset, would have bypassed the need for population-averaged AIF. This both simplifies processing and potentially spares the introduction of bias that can occur when AIFs derived from one study are applied to another. Additionally, the percentage change in $K^{trans}$ and $k_{ep}$ from all models is in line with previous literature (large decrease from baseline in CRs, no change to slight increase in non-responders)[44–46]. On the other hand, only percentage changes in $v_e$ from the RRM agree with previous reports, with complete responders showing static to slight decrease and non-responders showing significant increases[46]. Note that both the constrained fitting approach and RRIFT technique brought estimated parameter percentage change values into the expected ranges based on the previous analysis of this data[40], indicating their value.

The quantitative results acquired using the AIF-driven models were consistent with the previous quantitative analysis conducted on this dataset[40], which featured a multi-site analysis of single-site data. The median TM parameter estimates showed good agreement with those found in the previous publication: for example, on the first visit, median $K^{trans}$ estimates per patient ranged from 0.009–0.14 $min^{-1}$ in their study and our median values ranged from 0.02-0.09 $min^{-1}$. ETM estimates in this study were systematically slightly lower but still in relative agreement; first visit median $K^{trans}$ estimates per patient ranged over 0.019–0.15 $min^{-1}$ and our median values were approximately 0.01–0.05 $min^{-1}$. Seeing as the previous publication found large site-to-site variations between implementations of the same model, it is unsurprising that our implementation of the TM and ETM also produced different results. That said, the variation in our parameter estimates relative to those of Huang *et al.* remained within the range of variation across models and can likely be attributed to the same methodological sources they identified: AIF downsampling strategy, fitting method, assumed physiological/MR constants (for example, hematocrit, blood $T_1$, and bolus arrival time). A relatively high variance in $v_p$ estimates was also found among the different ETM implementations at the various test sites, which the authors attribute to the low temporal resolution of the dataset. Quantitative maps in this study showed similar heterogeneity to that reported by Huang *et al.*, owing to tumour heterogeneity. Our analysis of the percentage change of mean tumour $K^{trans}$ and $k_{ep}$ also showed good agreement with their reported findings among the TM and ETM implementations: for example, our $K^{trans}$ percent changes were between -65% to 30% for nCRs and -42% to -80% for CRs, compared with approximately -100% to 150% and -20% to -80%, respectively, in their study.

The differing behaviour between the CERRM and ETM estimates of $K^{trans}$ and $v_p$ merits attention. Previous simulation and *in-vivo* (glioblastoma) evaluation of both models showed that with reduced temporal sampling (via retrospective downsampling, in the case of the *in-vivo* data), the ETM began to underestimate $K^{trans}$ while the CERRM remained largely the same until much higher downsampling[32]. For $v_p$, both models appeared to overestimate, though the deviation in ETM estimates were much larger and appeared at a higher temporal resolution than the CERRM (ETM deviations began at approximately 10 s compared to 32 s for CERRM)[32]. With these previously-reported trends in mind, it is reasonable to assume that a similar effect is occurring in our findings due to the low temporal resolution, and that the true $K^{trans}$ and $v_p$ value lies closer to the estimates made by the CERRM due to our temporal resolution of 16-18 s. Studies using a dynamic phantom or more complex simulations would be useful to give a clearer picture on how estimates change in the case of breast imaging.

The conclusions made from this study are somewhat limited by the small size of this dataset. Additionally, the dataset is comprised of imaging conducted prior to 2013. Since this time, significant steps have been made in both hardware and acquisition methods which would likely allow for improved image quality and/or temporal resolution. To help guide the design of DCE-MRI studies, the Quantitative Imaging Biomarkers Alliance has produced recommendations for acquisition parameters based on the tissue under study [47]. Thus, for a more comprehensive analysis on the application of RRMs to breast cancer studies, use of a larger and more recent dataset would be prudent. During the course of this project one such dataset has been published, and could be considered in a future study of pharmacokinetic model performance [48].

A possible source of uncertainty in the methodology of this study was in the choice of reference regions. As stated, the first-choice reference region was the ipsilateral pectoralis major, which should in theory share the most similar AIF with the tumour region. In cases where the lesion infiltrated the ipsilateral pectoralis major, the contralateral pectoralis major was used, as in prior literature[8]. Finally, in some cases chest motion was too severe to accurately sample the pectoralis major, so serratus anterior was used. To our knowledge, this has not been used as a reference region in any other analysis of breast tumours with quantitative DCE-MRI; however, to help inform this decision, analysis was performed on one patient using the ipsilateral pectoralis major and anterior serratus, and results were found to be quite comparable. Regardless, selection of different muscles as reference regions may introduce some undesirable inter-patient variability that could affect the scaling of perfusion estimates and obscure patient-to-patient variations, since the tracer input to these muscles will not be identical to each other or the tumour region. This is not expected to be a limiting factor however, as AIF-reliant methods are often forced to use an anatomically distant vessel for AIF sampling, or a population-AIF, both of which can introduce erroneous inter-patient variability.

Another limitation was that this analysis was conducted solely with the population-averaged AIF provided with the dataset. In future work it would be ideal to include a comparison of all models, using both an image-derived and a population-averaged AIF, to observe how patient-to-patient variations change between models. Such a comparison however would likely favour the RRMs: all models would gain more patient-specific information, however the AIF-reliant models would likely see worse performance due to the poor resolution of the first pass AIF kinetics[21,24,25]. RRMs may be more robust to this, since most patient-specific information is contained in the first pass dynamics which is acquired from the reference region, while the AIF-tail (which provides the RRIFT scaling factor) is more constant across patients[20,21].

## Conclusion

This study showed that reference region models, which avoid the need for knowledge of first-pass AIF dynamics, can produce similar pharmacokinetic estimates of breast cancer tumours as traditional models using a high-resolution population-averaged AIF. Achieving similar quantification with less reliance on the AIF provides several advantages. First, this improves confidence in quantitative analysis at low-temporal resolutions where the AIF is poorly characterized, enabling more widespread use of RRMs in applications where temporal resolution must be sacrificed like breast imaging. Additionally, the positive prognostic ability of the relative RRMs (which require no AIF information at all) could allow for reliable, patient-specific quantification in cases where AIF sampling is not possible or in applications where no suitable literature-based AIF exists. Further investigation is required in assessing the differences in distribution of pharmacokinetic parameter estimates between models.

## Acknowledgments

Funding for this work was provided by a Discovery Grant from the Natural Science and Engineering Research Council (NSERC) of Canada, the Foundation of the Research Institute of the McGill University Health Centre, and NSERC Canada Graduate Scholarship Masters and *Fonds de recherche du Québec–Santé* Masters awards (CN). *In vivo* data was obtained from The Cancer Imaging Archive QIN Breast DCE-MRI collection. The authors acknowledge Dr. Jorge Campos for helpful discussions.

## Data availability statement

The code used in this study is available online at https://gitlab.com/MPUmri/breastRRMevaluation.

**Supplementary Information**

This Supplementary Information document is related to the manuscript:

*Validation of reference region modelling techniques in quantitative dynamic contrast-enhanced magnetic resonance imaging of breast cancer*

Christian Notte, Zaki Ahmed, and Ives R. Levesque

**Table S.1: Concordance correlation coefficients between each model, from the first imaging visit.**

| | | TM | ETM | CLRRM | CERRM |
|---|---|---|---|---|---|
| $K^{trans}$ | TM | - | 0.953 | 0.812 | 0.589 |
| | CLRRM | 0.812 | 0.734 | - | 0.749 |
| | CERRM | 0.589 | 0.554 | 0.749 | - |
| | | | | | |
| $v_e$ | TM | - | 0.952 | 0.913 | 0.846 |
| | CLRRM | 0.913 | 0.867 | - | 0.921 |
| | CERRM | 0.846 | 0.815 | 0.921 | - |
| | | | | | |
| $v_p$ | CERRM | - | 0.568 | - | - |

**Table S.2: Pearson correlation coefficient between each model, from the first imaging visit.**

| | | TM | ETM | CLRRM | CERRM |
|---|---|---|---|---|---|
| $K^{trans}$ | TM | - | 0.957 | 0.874 | 0.818 |
| | CLRRM | 0.874 | 0.810 | - | 0.809 |
| | CERRM | 0.818 | 0.813 | 0.809 | - |
| | | | | | |
| $v_e$ | TM | - | 0.953 | 0.947 | 0.879 |
| | CLRRM | 0.947 | 0.905 | - | 0.921 |
| | CERRM | 0.879 | 0.855 | 0.921 | - |
| | | | | | |
| $v_p$ | CERRM | - | 0.684 | - | - |

**Table S.3: Concordance correlation coefficient between each model, from the second imaging visit.**

| | | TM | ETM | CLRRM | CERRM |
|---|---|---|---|---|---|
| $K^{trans}$ | TM | - | 0.900 | 0.652 | 0.556 |
| | CLRRM | 0.652 | 0.505 | - | 0.832 |
| | CERRM | 0.556 | 0.488 | 0.832 | - |
| | | | | | |
| $v_e$ | TM | - | 0.922 | 0.787 | 0.745 |
| | CLRRM | 0.787 | 0.768 | - | 0.889 |
| | CERRM | 0.745 | 0.721 | 0.889 | - |
| | | | | | |
| $v_p$ | CERRM | - | 0.594 | - | - |

**Table S.4: Pearson correlation coefficient between each model, from the second imaging visit.**

| | | TM | ETM | CLRRM | CERRM |
|---|---|---|---|---|---|
| $K^{trans}$ | TM | - | 0.919 | 0.878 | 0.779 |
| | CLRRM | 0.878 | 0.775 | - | 0.836 |
| | CERRM | 0.779 | 0.794 | 0.836 | - |
| | | | | | |
| $v_e$ | TM | N/A | 0.925 | 0.865 | 0.831 |
| | CLRRM | 0.865 | 0.841 | - | 0.890 |
| | CERRM | 0.831 | 0.815 | 0.890 | - |
| | | | | | |
| $v_p$ | CERRM | - | 0.694 | - | - |

**Fig S.1: Box-and-whisker plots of the pharmacokinetic model RMSE, for each patient, prior to (A) and after (B) the first round of neoadjuvant chemotherapy made by each pharmacokinetic model. The box indicates the interquartile range, with the horizontal line contained within indicating the median value. The whiskers indicate the minimum and maximum values.**

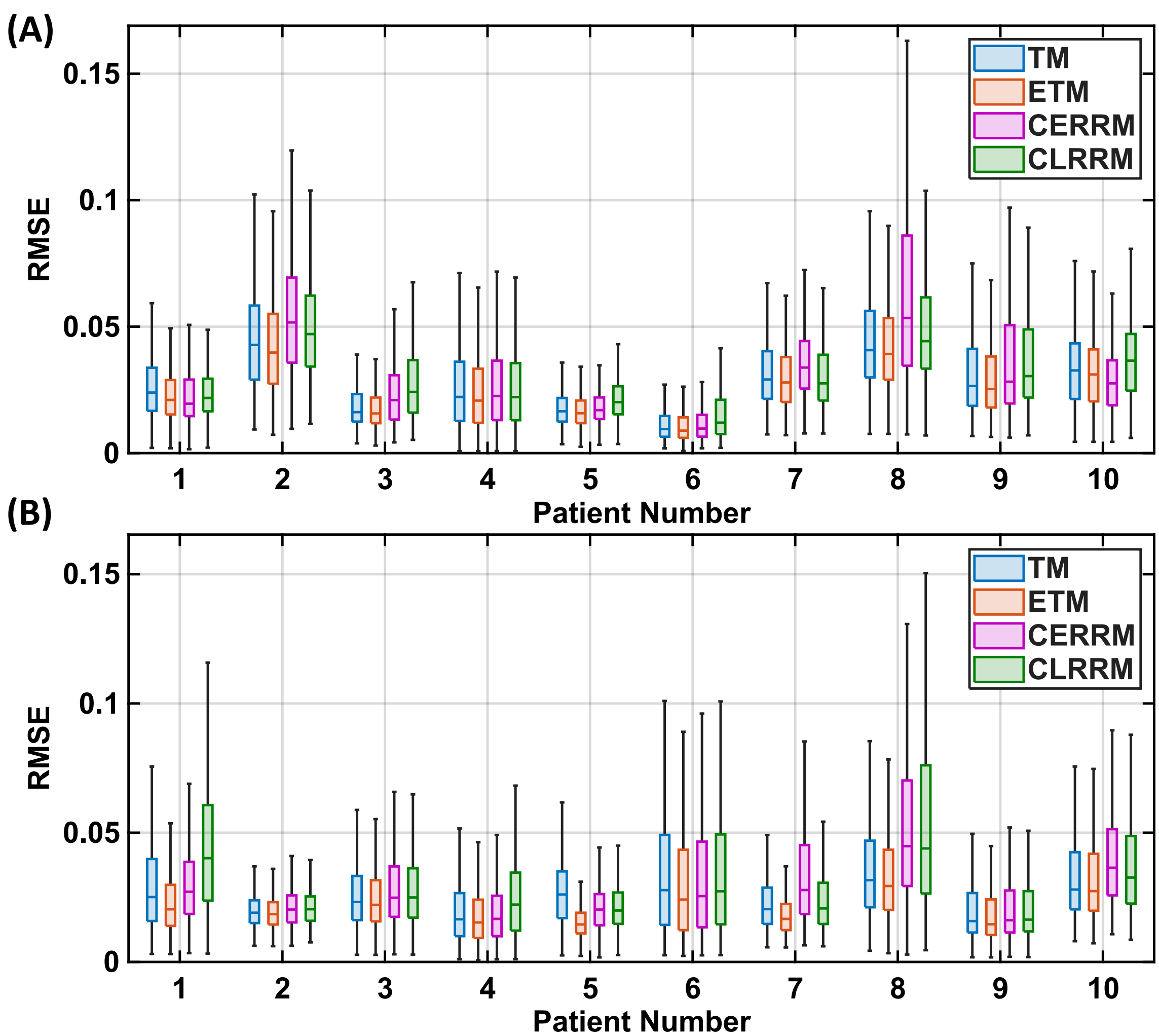

**Figure S.2: Box-and-whisker plots of $K^{trans}$ (A) and $v_p$ (B) estimates obtained from each pharmacokinetic model, for each patient after the first round of neoadjuvant chemotherapy. The box indicates the interquartile range, with the horizontal line contained within indicating the median value. The whiskers indicate the minimum and maximum values. A significant reduction in the spread of values can be seen among the complete responders (patient 2, 3, and 9).**

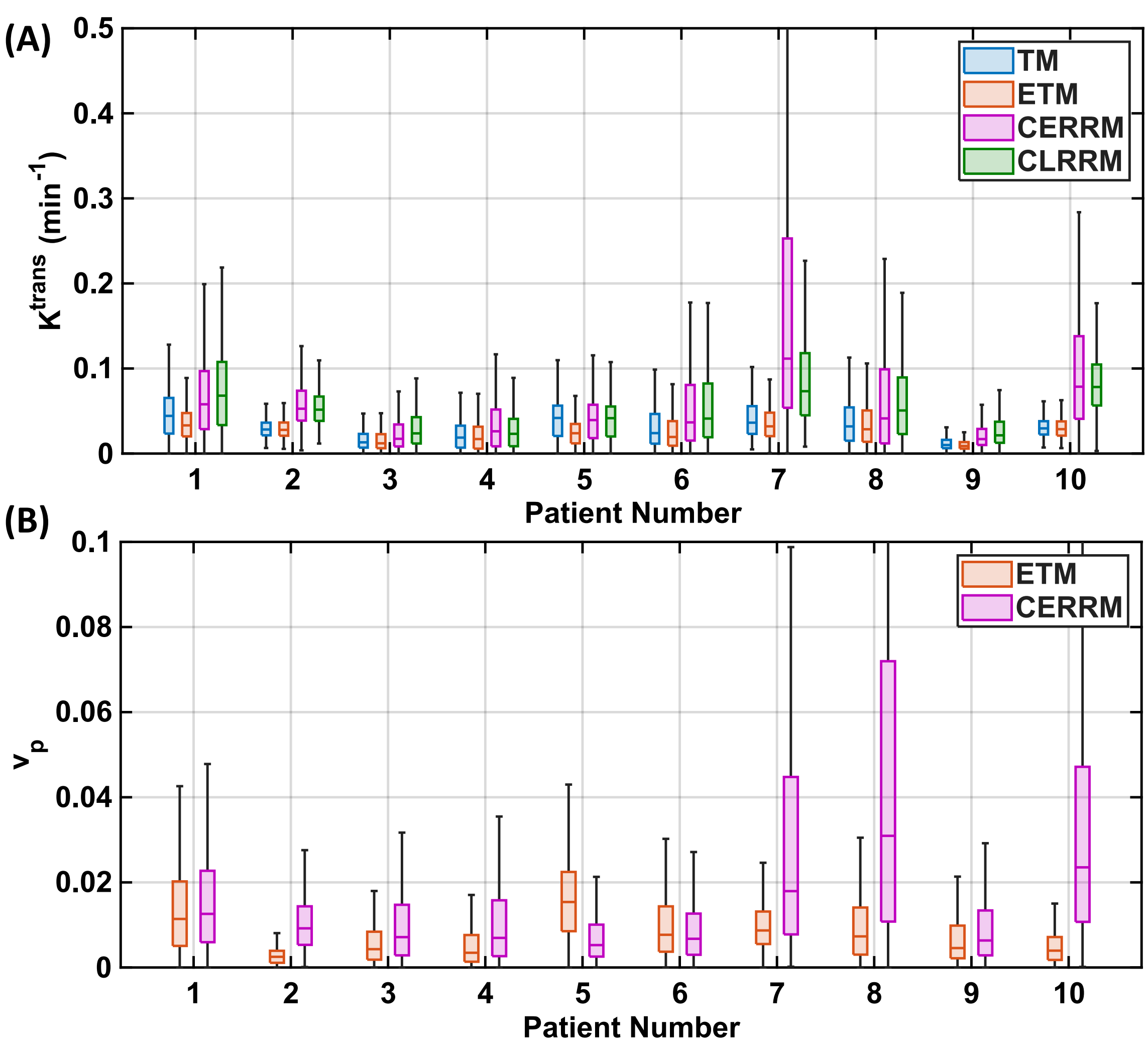